\documentclass[conference]{IEEEtran}
\IEEEoverridecommandlockouts
\usepackage{cite}
\usepackage{amsmath,amssymb,amsfonts}
\usepackage{algorithmic}
\usepackage{graphicx}
\usepackage{textcomp}
\usepackage{xcolor}
\usepackage{bm}
\usepackage{url}
\usepackage{comment}
\usepackage{mathtools}
\def\BibTeX{{\rm B\kern-.05em{\sc i\kern-.025em b}\kern-.08em
    T\kern-.1667em\lower.7ex\hbox{E}\kern-.125emX}}
\newtheorem{example}{Example}
\newtheorem{theorem}{Theorem}
\newtheorem{definition}{Definition}
\newtheorem{lemma}{Lemma}
\newtheorem{prop}{Proposition}
\newtheorem{assumption}{Assumption}

\allowdisplaybreaks[4]
\begin{document}

\title{Stochastic 2D Signal Generative Model with Wavelet Packets Basis Regarded as a Random Variable and Bayes Optimal Processing
{\thanks{This work was supported by JSPS KAKENHI Grant Numbers JP17K06446 and JP19K04914}}
}

\author{\IEEEauthorblockN{Ryohei Oka}
\IEEEauthorblockA{\textit{Dept. Pure and Applied Mathematics} \\
\textit{Waseda University}\\
Okubo, Shinjuku-ku, Tokyo, Japan \\
ryohei1207@toki.waseda.jp}
\and
\IEEEauthorblockN{Yuta Nakahara}
\IEEEauthorblockA{\textit{Center for Data Science} \\
\textit{Waseda University}\\
Nishiwaseda, Shinjuku-ku, Tokyo, Japan \\
yuta.nakahara@aoni.waseda.jp}
\and
\IEEEauthorblockN{Toshiyasu Matsushima}
\IEEEauthorblockA{\textit{Dept. Pure and Applied Mathematics} \\
\textit{Waseda University}\\
Okubo, Shinjuku-ku, Tokyo, Japan \\
toshimat@waseda.jp}
}

\maketitle

\begin{abstract}
This study deals with two-dimensional (2D) signal processing using the wavelet packet transform. When the basis is unknown the candidate of basis increases in exponential order with respect to the signal size. Previous studies do not consider the basis as a random variable. Therefore, the cost function needs to be used to select a basis. However, this method is often a heuristic and a greedy search because it is impossible to search all the candidates for a huge number of bases. Therefore, it is difficult to evaluate the entire signal processing under a criterion, and also it does not always guarantee the optimality of the entire signal processing. In this study, we propose a stochastic generative 
model in which the basis is regarded as a random variable. This makes it possible to evaluate entire signal processing under a unified criterion i.e. Bayes criterion. Moreover, we can derive an optimal signal processing scheme that achieves the theoretical limit. This derived scheme shows that all the bases should be combined according to the posterior instead of selecting a single basis. Although exponential order calculations are required for this scheme, we have derived a recursive algorithm for this scheme, which successfully reduces the computational complexity from the exponential order to the polynomial order.
\end{abstract}


\section{Introduction}
This study deals with two-dimensional (2D) signal processing using the wavelet packet transform. Specifically, We perform 2D signal processing based on statistical decision theory (see e.g. \cite{berger}) with Bayes risk function as the evaluation criterion.
 
Wavelet packet transform has been applied in various fields, and a lot of research has been done in recent years. For example, denoising (see e.g. \cite{denoising_study1,denoising_study2}), compression (see e.g. \cite{compression_study1}) , classification (see e.g. \cite{classification_study1}), and so on. In any case, they have difficulty in finding an optimal basis because the candidate of basis increases in exponential order with respect to the size of the signal.
 
Most previous studies do not consider the basis as a random variable. Therefore, some cost functions, for example, such as Shannon entropy \cite{denoising_study1}, need to be used to search and select a basis. However, this method is often a heuristic and a greedy search because it is impossible to search all the candidates for a huge number of bases. Therefore, it is difficult to evaluate the entire signal processing under a criterion, and also it does not always guarantee the optimality of the entire signal processing.

In this study, we propose a stochastic generative model in which the basis is regarded as a random variable. In other words, we consider Bayesian modeling of the basis. This makes it possible to evaluate entire signal processing under a criterion called Bayes risk function in the framework of statistical decision theory\cite{berger}. Moreover, we can derive an optimal signal processing scheme that achieves the theoretical limit by using well-known theorems in the framework of statistical decision theory. In the derived scheme, all the bases are weighted by the posterior probability. This shows that any single basis should not be chosen under the Bayes criterion.

In this study, we use this stochastic generative model for denoising. However, there is a problem with the scheme that achieves the theoretical limit. The problem is that the amount of computation increases exponentially with respect to the size of the signal. 
To solve this problem, we have derived a recursive algorithm that utilizes the property of prior distribution assumed for the basis. This algorithm successfully reduces the computational complexity from the exponential order to the polynomial order without loss of optimality. 

In section \ref{experiment} we conduct two numerical experiments about the derived algorithm. The first is an experiment on the posterior distribution of the basis. In this experiment, we confirm that the posterior probability of the true basis is high. The second is an experiment to see the value of Bayes risk function. In this experiment, we compare the value of the Bayes risk function of the proposed method with that of the denoising method under some fixed wavelet packet bases to confirm the effectiveness of the proposed methods.
\section{Proposed model}
In this section, we define Walsh wavelet packets basis (see e.g. \cite{text}) and propose a stochastic model. In the following, we assume the 2D signal size is $L\times L=2^{d_{\max}}\times 2^{d_{\max}}(d_{\max}\in \mathbb{N}\cup \{0\})$.
\subsection{2D wavelet packets \cite{text}}
\begin{definition}
($w_{i,j}(n)$)\\
 The function $w_{i,j}:\mathbb{Z}\to \mathbb{R}$ is defined as follows.
\begin{align}
w_{0,0}(n)&\coloneqq
\begin{cases}
1&(n=0)\\
0&(\text{otherwise}),
\end{cases}
\\
w_{i+1,2j}(n)&\coloneqq2^{-1/2}w_{i,j}(n)+2^{-1/2}w_{i,j}(n-2^i),\\
w_{i+1,2j+1}(n)&\coloneqq2^{-1/2}w_{i,j}(n)-2^{-1/2}w_{i,j}(n-2^i),
\end{align}
where $i\in \{0,1,\cdots,d_{\max}\}$ and $j\in \{0,1,\cdots,2^i-1\}$.
\end{definition}

\begin{definition}
($w_{i,j_0,j_1}(n_0,n_1)$)\\
The function $w_{i,j_0,j_1}:\mathbb{Z} \times \mathbb{Z}\to \mathbb{R}$ is defined as follows.
\begin{align}
w_{i,j_0,j_1}(n_0,n_1)&\coloneqq w_{i,j_0}(n_0)w_{i,j_1}(n_1),
\end{align}
where $i\in \{0,1,\cdots,d_{\max}\}$ and $j_0,j_1\in \{0,1,\cdots,2^i-1\}.$
\end{definition}

\begin{definition}
($W_{i,j_0,j_1,k_0,k_1}$)\\
The matrix $W_{i,j_0,j_1.k_0,k_1}\in \mathbb{R}^{L\times L}$ is defined as follows. 
\begin{align}
&W_{i,j_0,j_1,k_0,k_1} \nonumber\\
&\coloneqq(w_{i,j_0,j_1}(n_0-2^ik_0,n_1-2^ik_1))_{n_0,n_1\in \{0,1,\cdots,L\}},
\end{align}
where $k_0,k_1\in \{0,1,\cdots,L/2^i-1\}.$
\end{definition}

\begin{definition}
($\mathcal{W}_{i,j_0,j_1}\subseteq \mathbb{R}^{L\times L}$)\\
Let $\mathcal{W}_{i,j_0,j_1}\subseteq \mathbb{R}^{L\times L}$ be the space of  2D signals that consist of a linear combination of $\{W_{i,j_0,j_1,k_0,k_1}\}_{k_0,k_1\in \{0,1,\cdots,L/2^i-1\}}$.
\end{definition}

\begin{definition}
($\bm{w}_{i,j_0,j_1,k_0,k_1}$)\\
Let $\bm{w}_{i,j_0,j_1,k_0,k_1}$ be the vertical vector whose components are those of $W_{i,j_0,j_1,k_0,k_1}$ rearranged in raster scan order.
\end{definition}

Under the above definition, the following property holds.
\begin{prop}
($\mathcal{W}_{0,0,0}$)\\
The following relationship holds
\begin{align}
\mathcal{W}_{0,0,0}&=\mathbb{R}^{L\times L}.
\end{align}
\end{prop}

\begin{prop}
(Orthonormal basis of $\mathcal{W}_{i,j_0,j_1}$)\\
$\{W_{i,j_0,j_1,k_0,k_1}\}_{k_0,k_1\in\{0,1,\cdots,L/2^i-1\}}$ forms an orthonormal basis for $\mathcal{W}_{i,j_0,j_1}$.
\end{prop}

\begin{prop}
(Decomposition of space)\\\
The following relationship holds.
\begin{align}
\mathcal{W}_{i,j_0,j_1}=&\hspace{0.1cm} \mathcal{W}_{i+1,2j_0,2j_1}\oplus \mathcal{W}_{i+1,2j_0,2j_1+1}
\nonumber \\
&\oplus \mathcal{W}_{i+1,2j_0+1,2j_1}\oplus \mathcal{W}_{i+1,2j_0+1,2j_1+1}.
\end{align}
\end{prop}
Note that $\oplus$ denotes the orthogonal direct sum.

From the aforementioned properties, the entire space of 2D signals can be represented as an orthogonal direct sum of subspaces corresponding to the leaf nodes of the full quadtree\footnote{All nodes have 4 child nodes or no child node.} with $\mathcal{W}_{0,0,0}$ as the root node. The basis of each subspace can be used to construct an orthonormal basis.

\begin{example}
(Full quadtree)\\
When the relation $\mathcal{W}_{0,0,0}=(\mathcal{W}_{2,0,0}\oplus \mathcal{W}_{2,0,1}\oplus \mathcal{W}_{2,1,0}\oplus \mathcal{W}_{2,1,1})\oplus \mathcal{W}_{1,0,1}\oplus\mathcal{W}_{1,1,0}\oplus\mathcal{W}_{1,1,1}$ holds, the diagram of the full quadtree is at Fig. \ref{ex_tree}. The gray area represents leaf nodes. Notably, the basis of 2D-DWT corresponds to the full quadtree created by extending branches only in the direction $j_0=j_1=0$.

\begin{figure}[tbp]
\centering
\includegraphics[width=6cm,height=3cm]{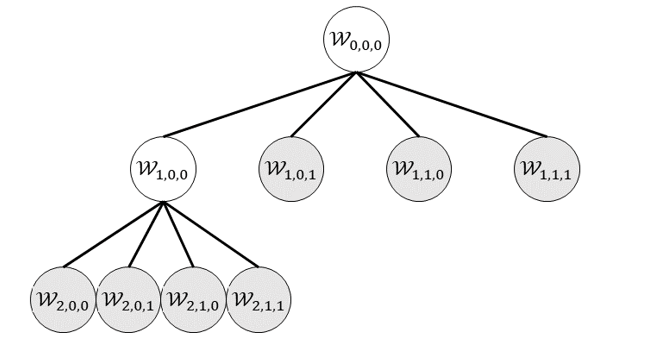}
\caption{example for quadtree under $d_{\max}=2$}
\label{ex_tree}
\end{figure}
\end{example}

\subsection{2D wavelet packets basis matrix}
\begin{definition}
($\mathcal{S},\mathcal{L},\mathcal{I},s_{\text{r}}$)\\
Let $\mathcal{S}$ be the set of nodes in a complete quadtree\footnote{All leaf nodes have the same depth.} of depth $d_{\max}$ and  $\mathcal{L}\subset \mathcal{S}$ be the set of its leaf nodes and $\mathcal{I}\subset \mathcal{S}$ be the set of its inner nodes. Let $s_{\text{r}}$ be  the root node.
\end{definition}

\begin{definition}
($\mathcal{M},\mathcal{L}^m,\mathcal{I}^m$)\\
Let $\mathcal{M}$ be the set of all full quadtrees on $\mathcal{S}$, which contain $s_{\text{r}}$ and whose depth is less than or equal to $d_{\max}$. Let $\mathcal{L}^m \subset \mathcal{S}$ be the set of leaf nodes in $m \in \mathcal{M}$ and $\mathcal{I}^m \subset \mathcal{S}$ be the set of inner nodes in $m \in \mathcal{M}$.
\end{definition}

\begin{definition}
($W^m$)\\
Let $W^m\in \mathbb{R}^{L^2 \times L^2}$ be a matrix created by collecting the basis of the subspace corresponding to the leaf nodes in $m \in \mathcal{M}$. The matrix $W^m\in \mathbb{R}^{L^2 \times L^2}$ is called the basis matrix in this paper.
\end{definition}
\begin{definition}
($W_s$)\\
Let $W_s \in \mathbb{R}^{L^2 \times L^2}$ be the basis at $s\in \mathcal{S}$ which is appropriately complemented by 0 vectors so that the following relation holds.
\begin{align}
W^m&=\sum_{s\in \mathcal{L}^m}W_s.
\end{align}
\end{definition}

\begin{example}
(The basis matrix corresponding to the quadtree $m\in \mathcal{M}$ in Figure \ref{ex_tree})
\begin{align}
W^m
&=
\begin{pmatrix}
\bm{w}_{1,0,0,0,0}^{\top}\\
\bm{w}_{1,0,0,0,1}^{\top}\\
\bm{w}_{1,0,0,1,0}^{\top}\\
\bm{w}_{1,0,0,1,1}^{\top}\\
\bm{0}_{12\times 16}
\end{pmatrix}
+
\begin{pmatrix}
\bm{0}_{4\times 16}\\
\bm{w}_{1,0,1,0,0}^{\top}\\
\bm{w}_{1,0,1,0,1}^{\top}\\
\bm{w}_{1,0,1,1,0}^{\top}\\
\bm{w}_{1,0,1,1,1}^{\top}\\
\bm{0}_{8\times 16}
\end{pmatrix}
+
\begin{pmatrix}
\bm{0}_{8\times 16}\\
\bm{w}_{1,1,0,0,0}^{\top}\\
\bm{w}_{1,1,0,0,1}^{\top}\\
\bm{w}_{1,1,0,1,0}^{\top}\\
\bm{w}_{1,1,0,1,1}^{\top}\\
\bm{0}_{4\times 16}\\
\end{pmatrix}
\nonumber\\
&
+
\begin{pmatrix}
\bm{0}_{12\times 16}\\
\bm{w}_{2,2,2,0,0}^{\top}\\
\bm{0}_{3\times 16}\\
\end{pmatrix}
+
\begin{pmatrix}
\bm{0}_{13\times 16}\\
\bm{w}_{2,2,3,0,0}^{\top}\\
\bm{0}_{2\times 16}\\
\end{pmatrix}
+
\begin{pmatrix}
\bm{0}_{14\times 16}\\
\bm{w}_{2,3,2,0,0}^{\top}\\
\bm{0}_{1\times 16}\\
\end{pmatrix}
\nonumber\\
&+
\begin{pmatrix}
\bm{0}_{15\times 16}\\
\bm{w}_{2,3,3,0,0}^{\top}\\
\end{pmatrix}
.
\end{align}
where $\bm{0}_{k\times l}$ is a $k\times l$ zero matrix.
\end{example}

The basis matrix $W^m$ is orthonormal because it consists of an orthonormal basis.
\subsection{Stochastic generative model}
Herein, we describe the stochastic 2D signal model.
\begin{definition}
($\bm{X},\bm{x}$)\\
Let $\bm{X}$ be a random variable vector on $\mathbb{R}^{L^2\times 1}$ representing 2D signal and $\bm{x}\in \mathbb{R}^{L^2\times 1}$ be its realization. $\bm{x}\in\mathbb{R}^{L^2\times 1}$ is a vertical vector that is given by sorting $L\times L$ size 2D signal in raster scan order.  2D-DWPT $\bm{\theta}$ defined below also has a similar structure.
\end{definition}

\begin{definition}
($\bm{\theta}$)\\
Let $\bm{\theta}$ be a random variable vector on $\mathbb{R}^{L^2\times 1}$ representing the 2D-DWPT. Its realization is similarly written as $\bm{\theta}\in\mathbb{R}^{L^2\times 1}$.
\end{definition}

The following relationship holds between $\bm{x}$ and $\bm{\theta}$.
\begin{align}
\bm{\theta}&=W^m\bm{x},\\\
\bm{x}&=(W^m)^{\top}\bm{\theta}.
\end{align}

In our model, the quadtree $m \in \mathcal{M}$ and 2D-DWPT $\bm{\theta}$ are generated by assumed prior distributions, and the 2D signal $\bm{x}$ is generated by a linear transformation of $\bm{\theta}$.

\section{Application to Bayesian signal processing}\label{problem}
Under the generative probability model discussed in the previous section, various signal processing problems can be considered depending on how the input signal is observed and the nature of the desired output signal. In the Bayesian decision theory\cite{berger}, these are represented by designing the domain and the range of decision function as well as the loss function.
In this section, we formulate the simplest denoising problem based on the Bayesian decision theory and find the optimal denoising under the Bayesian criterion.

\subsection{Problem}
The following additive noises are considered in this paper.
\begin{assumption}\label{noise_assumption}
(additive noise)
\begin{align}
\bm{Y}\coloneqq\bm{X}+\bm{\epsilon}.
\end{align}
We assume $\bm{Y}$ is a vector of random variables on $\mathbb{R}^{L^2\times 1}$ representing the noisy 2D signal, $\bm{X}$ is a random variable representing the original 2D signal, and $\bm{\epsilon}\sim \mathcal{N}(\bm{\epsilon }|\bm{0}_{L^2\times 1},\sigma^2_{\bm{\epsilon}}I)$ ($\sigma^2_{\bm{\epsilon}}\in \mathbb{R}$ is a known hyperparameter). In the following, let $\bm{y}\in \mathbb{R}^{L^2\times 1}$ be the realization of $\bm{Y}$. Our goal in Sections \ref{problem} and \ref{experiment} is to estimate $\bm{x}$ from $\bm{y}$.
\end{assumption}

We now define the decision function $\delta$.
\begin{definition}
(Decision function $\delta$)\\
The function $\delta:\mathbb{R}^{L^2\times 1}\to \mathbb{R}^{L^2\times 1}$ is called the decision function. The input of the decision function $\delta$ is the noisy 2D signal $\bm{y}$, and the output is an estimation for the original 2D signal $\hat{\bm{x}}$.
\end{definition}

We define the Bayes optimal decision function $\delta^{\ast}$ as the function that minimizes the Bayes risk function BR($\delta$) based on mean-square error loss in the 2D signal domain.
\begin{definition}
(Bayes risk function $BR(\delta)$)\\\
The Bayes risk function $BR:\Delta \to \mathbb{R}$ ($\Delta$ is the space of decision functions $\delta$) is defined as below. 
\begin{align}
BR(\delta)\coloneqq\sum_{m\in \mathcal{M}} \int \int &\frac{1}{L^2}\|\bm{x}-\delta(\bm{y})\|^2\nonumber \\
&\times p(\bm{y}|\bm{x},m)p(\bm{x}|m)p(m) \mathrm{d}\bm{y} \mathrm{d}\bm{x}.
\end{align}
\end{definition}

\subsection{Solution to the problem}
According to the Bayesian decision theory (see e.g. \cite{berger}), the following holds.
\begin{theorem}
(Bayes optimal decision function $\delta^{\ast}$)\\
Bayes optimal decision function $\delta^{\ast}$ is given by
\begin{align}\label{Bayes_optimal}
\delta^{\ast}(\bm{y})&=\sum_{m\in \mathcal{M}}p(m|\bm{y})\int \bm{x} p(\bm{x}|m,\bm{y})\mathrm{d}\bm{x}.
\end{align}
\end{theorem}
There are two problems in the calculation of the right-hand side of the equation (\ref{Bayes_optimal}).
First, the integral of $\bm{x}$ does not generally have a closed-form expression. Second, the computation complexity of summation with respect to $m \in \mathcal{M}$ increases exponentially with respect to the size of the 2D signal. However, using an appropriate prior distribution of $\bm{\theta}$ and the prior distribution of the quadtree model $m\in \mathcal{M}$, the computational complexity can be reduced to polynomial order while maintaining optimality.

\subsection{Efficient algorithm for the solution}
Let $\bm{\mu}^m\in \mathbb{R}^{L^2\times 1}$ and $\sigma^2\in \mathbb{R}$ be a known hyperparameters. $\bm{\mu}^m$ is constructed in the same way as $W^m$, where $\bm{\mu}^m=\sum_{s\in \mathcal{L}^m}\bm{\mu}_s\hspace{0.1cm}(\bm{\mu}_s\in \mathbb{R}^{L^2\times 1}$ is also a known vector).
\begin{assumption}\label{prior_theta}
(2D-DWPT prior distribution)
\begin{align}
p(\bm{\theta}|m)&\coloneqq\mathcal{N}(\bm{\theta}|\bm{\mu}^m, \sigma^2 I),
\end{align}
where $I$ is the $L^2\times L^2$ identity matrix.
\end{assumption}
By linear transformation, we can show that the 2D signal $\bm{X}$ follows the normal distribution as below.
\begin{prop}\label{signal_distribution}
(Distribution of the 2D signal)\\
The distribution of the 2D signal $\bm{X}$ is given by
\begin{align}
p(\bm{x}|m)&=\mathcal{N}(\bm{x}|(W^m)^{\top}\bm{\mu}^m, \sigma^2I).
\end{align}
\end{prop}
\begin{assumption}\label{tree}
(Prior distribution of the quadtree model)
\begin{align}
p(m)&=\prod_{s\in \mathcal{L}^m}(1-g_{s}) \prod_{s'\in \mathcal{I}^m}g_{s'}\label{priorm}
\end{align}
where $g_{s}\in [0,1]$ is a known hyperparameter for any $s\in \mathcal{S}$, which satisfies $g_{s}=0$ for $s \in \mathcal{L}$. The $g_s$ represents the probability that node $s$ will extend a branch.
\end{assumption}

The aforementioned prior distribution (\ref{priorm}) was proposed in \cite{matsushima} to represent context trees and applied in \cite{110003313864} to represent the prior distribution of 1D wavelet packets trees. A mathematically rigorous proof of the following equation (\ref{prob}), their expected values, and posterior probability calculations is given in \cite{nakahara}. 
\begin{align}
\sum_{m\in \mathcal{M}}p(m)=1\label{prob}
\end{align}
There has not been any previous study that applied this prior distribution to 2D-DWPT trees to our best knowledge.  Hence, we applied this prior distribution to 2D-DWPT
trees for the first time.

Under the aforementioned assumptions, the following theorem can be derived.

\begin{theorem}\label{theorem_posterior}
(Posterior distribution)\\
The following equation holds.
\begin{align}
p(\bm{x}|m,\bm{y})&=\mathcal{N}(\bm{x}|\tilde{\bm{\mu}}^m,\tilde{\sigma}^2 I),\label{p(x|m,y)}\\
p(m|\bm{y})&=\prod_{s\in \mathcal{L}^m}(1-\tilde{g}_s)\prod_{s' \in \mathcal{I}^m}\tilde{g}_{s'}\label{p(m|y)},
\end{align}
where
\begin{align}
\tilde{\bm{\mu}}^m&\coloneqq \frac{\sigma^2}{\sigma^2+\sigma^2_{\bm{\epsilon}}} \bm{y}+\frac{\sigma^2_{\bm{\epsilon}}}{\sigma^2+\sigma^2_{\bm{\epsilon}}}(W^m)^{\top}\bm{\mu}^m,\\
\tilde{\sigma}^2&\coloneqq \frac{\sigma^2\sigma^2_{\bm{\epsilon}}}{\sigma^2+\sigma^2_{\bm{\epsilon}}},\\
\tilde{g}_s&\coloneqq
\begin{cases}
g_s&(s\in \mathcal{L})\\
\cfrac{g_s\prod_{s'\in \text{Ch}(s)}\tilde{\psi}_{s'}}{\tilde{\psi}_s}&(\text{otherwise}),
\end{cases}
\\
\tilde{\psi}_s&\coloneqq
\begin{cases}
\psi_s &(s\in \mathcal{L})\\
(1-g_s)\psi_s+g_s\prod_{s'\in \text{Ch}(s)}\tilde{\psi}_{s'}&(\text{otherwise}),
\end{cases}
\\
\ln \psi_s&\coloneqq\frac{1}{\sigma^2+\sigma^2_{\bm{\epsilon}}}\bigg(W_s\bm{y}-\frac{\bm{\mu}_s}{2}\bigg)^{\top}\bm{\mu}_s.
\end{align}

Note that $\text{Ch}(s)\subset \mathcal{S}$ is the set of child nodes of $s\in \mathcal{S}$.
\end{theorem}
The proof of this theorem is in Appendix \ref{proof_posterior}.\\
Using Theorem \ref{theorem_posterior}, the Bayes optimal decision function $\delta^{\ast}$ can be calculated as follows.

\begin{theorem}
(Bayes optimal decision function $\delta^{\ast}$)\\
The following equation holds
\begin{align}
\delta^{\ast}(\bm{y})&=\frac{\sigma^2}{\sigma^2+\sigma^2_{\bm{\epsilon}}}  \bm{y}+\frac{\sigma^2_{\bm{\epsilon}}}{\sigma^2+\sigma^2_{\bm{\epsilon}}} \sum_{m\in \mathcal{M}}p(m|\bm{y})(W^m)^{\top}\bm{\mu}^m.\label{Bayes}
\end{align}
\end{theorem}
The integral of $\bm{x}$ in (\ref{Bayes_optimal}) is solved in (\ref{Bayes}) but the computational complexity of the second term increases exponentially with the size of the 2D signal. However, using the recursive computation derived from the following theorem, the computational complexity can be reduced to polynomial order while maintaining Bayesian optimality.
\begin{theorem}\label{theorem_recursive}
(Recursive algorithm)\\
The following equation holds.
\begin{align}
r_{s_{\text{r}}}=\sum_{m\in \mathcal{M}}p(m|\bm{y})(W^m)^{\top}\bm{\mu}^m.
\end{align}
\begin{align}
r_s&\coloneqq
\begin{cases}
(1-\tilde{g}_s)W^{\top}_s\bm{\mu}_s&(s\in \mathcal{L})\\
(1-\tilde{g}_s)W^{\top}_s\bm{\mu}_s+\tilde{g}_s\sum_{s'\in \text{Ch}(s)}r_{s'}&(\text{otherwise}).
\end{cases}
\label{r_s}
\end{align}
\end{theorem}
The proof of this theorem is in Appendix \ref{proof_recursive}.
\section{Experiments}\label{experiment}
\subsection{Experiment 1 : Posterior $p(m|\bm{y})$}
The purpose of Experiment 1 is to quantitatively confirm that the posterior distribution $p(m|\bm{y})$ is properly computed.

\begin{figure}[tbp]
\centering
\includegraphics[height=5cm,width=8cm]{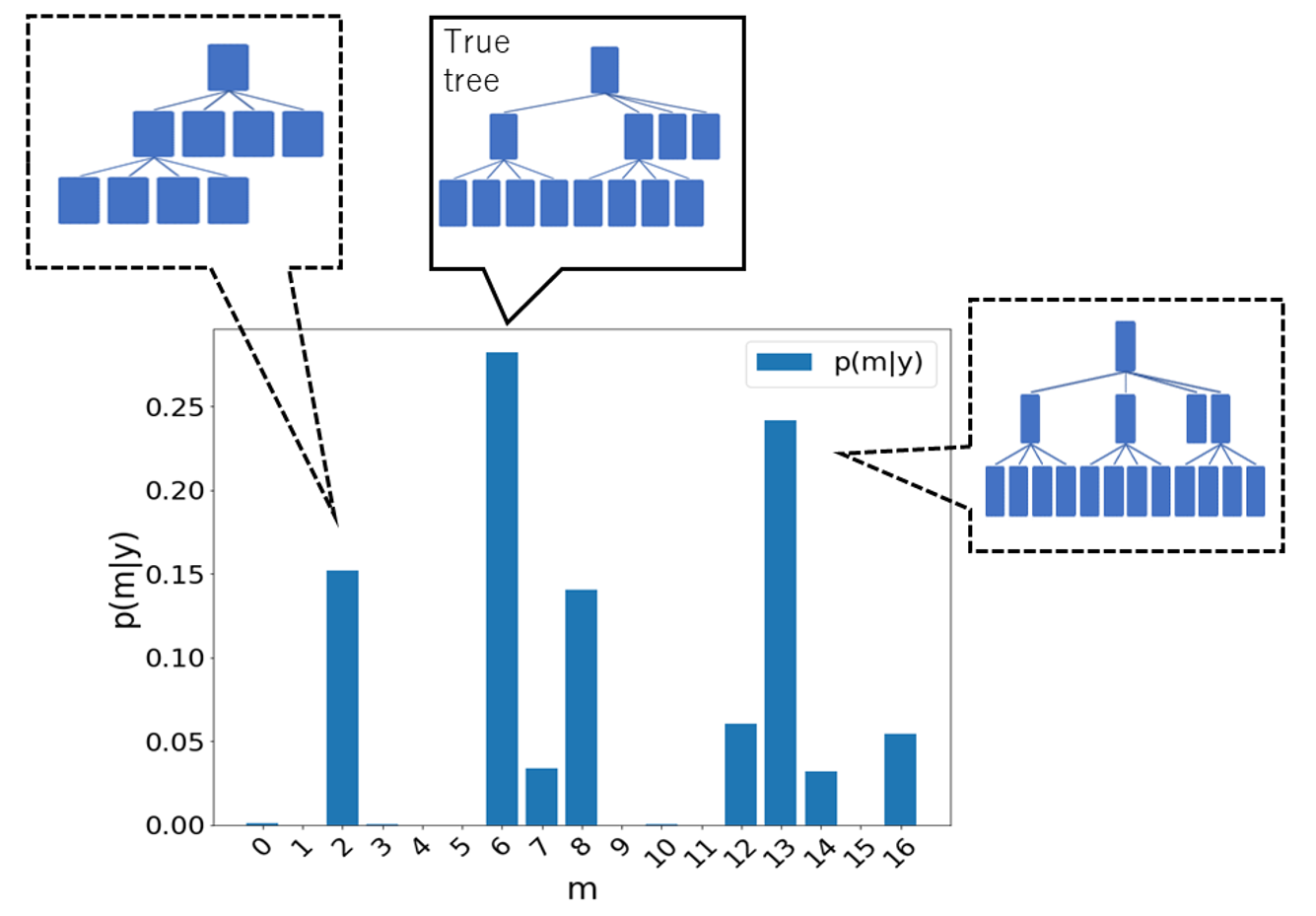}
\caption{$p(m|\bm{y})$ in Experiment 1}
\label{experiment1}
\end{figure}

The setting of Experiment 1 is as follows. 
\begin{itemize}
\item $d_{\max}=2$.
\item $g_s = 0.5$.
\item $\sigma^2 = 10$.
\item $\sigma^2_{\bm{\epsilon}}=4$.
\item $\bm{\mu}_s$ is calculated from resized images in \cite{data}. \footnote{Let $\bm{x}_1,\cdots,\bm{x}_{50000}$ denote the training images in \cite{data}. Non zero terms of $\bm{\mu}_s$ is calculated as $ \frac{1}{50000\times 4^{d_{\max}-i(s)}}\sum_{n=1}^{50000}f(W_s\bm{x}_n)$, where $f:\mathbb{R}^N \to \mathbb{R}$ is defined as $f((a_1,a_2,\cdots,a_N))\coloneqq \sum_{n=1}^{N}a_n $ and $i(s)$ is the depth of node $s$.}
\end{itemize}
This experiment is conducted by using the following procedure.
\begin{enumerate}
\item Set $m_0$ (6 in Fig.\ref{experiment1}).\label{ex2_genem}
\item Generate $\bm{x} \sim p(\bm{x}|m_0)$.\label{ex2_genex}
\item Generate $\bm{y} = \bm{x} + \bm{\epsilon}$.\label{ex2_geney}
\item Calculate $\tilde{g}_s$.\label{ex2_calp}
\item Repeat from step \ref{ex2_geney} to step \ref{ex2_calp} 50 times. \label{ex2_re1}
\item Repeat from step \ref{ex2_genex} to step \ref{ex2_re1} 50 times.\label{ex2_re2}
\end{enumerate}
The result of this experiment is in Fig. \ref{experiment1}.
Since $d_{\max}=2$, we have $|\mathcal{M}|=17$. In other words, there are 17 candidate trees in total. we give them indices from 0 to 16 in an arbitrary order. In this experiment, the sixth tree in this order is fixed as the true tree for data generation.

According to this result, we can confirm that the posterior probability of the sixth tree, which is the true tree, is the maximum value. Moreover, the tree with only one node expanded from the true tree has the second-largest posterior probability and the tree with only one node shrunk from the true tree has the third-largest posterior probability. Therefore, we confirmed our algorithm properly computed the posterior distribution by not only the theoretical proof but also the quantitative experiment.

\subsection{Experiment 2 : Value of Bayes risk function $BR(\delta)$}
The purpose of Experiment 2 is to confirm the effectiveness of the proposed method by calculating the value of the Bayes risk function.
We compare the proposed method with the estimated signals from five models $m_{i} \in \mathcal{M} \hspace{0.1cm}(i=1,2,3,4,5)$. $m_i$ is a perfect quadtree expanded to the depth $i$.

The estimated signals from these models $m_{i} \in \mathcal{M}$ are obtained by the following equations.
\begin{align}
\delta^{i}(\bm{y})&=\frac{\sigma^2}{\sigma^2+\sigma^2_{\bm{\epsilon}}}\bm{y}+\frac{\sigma^2_{\bm{\epsilon}}}{\sigma^2+\sigma^2_{\bm{\epsilon}}}(W^{m_{i}})^{\top}\bm{\mu}^{m_{i}}.
\end{align}

The setting of Experiment 2 is as follows. 
\begin{itemize}
\item $d_{\max}=5$.
\item $\sigma^2 = 10$.
\item $\sigma^2_{\bm{\epsilon}}=4$.
\item $\bm{\mu}_s$ is calculated from images in \cite{data}.
\end{itemize}
This experiment is conducted by using the following procedure.
\begin{enumerate}
\item Set $g_s \in \{0.1,0.2,0.3,0.4,0.5,0.6,0.7,0.8,0.9\}$.
\item Gernerate quadtree $m \sim p(m)$.\label{ex3_genem}
\item Generate $\bm{x}\sim p(\bm{x}|m)$.\label{ex3_genex}
\item Generate $\bm{y} = \bm{x}+\bm{\epsilon}$.\label{ex3_geney}
\item Calculate $\delta^{\ast}(\bm{y}), \delta^{i}(\bm{y})$.\label{ex3_caldelta}
\item Calculate mean-square error loss.\label{ex3_calloss}
\item Repeat from step \ref{ex3_geney} to step \ref{ex3_calloss} 10 times.\label{ex3_rep1}
\item Repeat from step \ref{ex3_genex} to step \ref{ex3_rep1} 10 times.\label{ex3_rep2}
\item Repeat from step \ref{ex3_genem} to step \ref{ex3_rep2} 30 times.
\end{enumerate}

\begin{figure}[tbp]
\centering
\includegraphics[height=5cm,width=8cm]{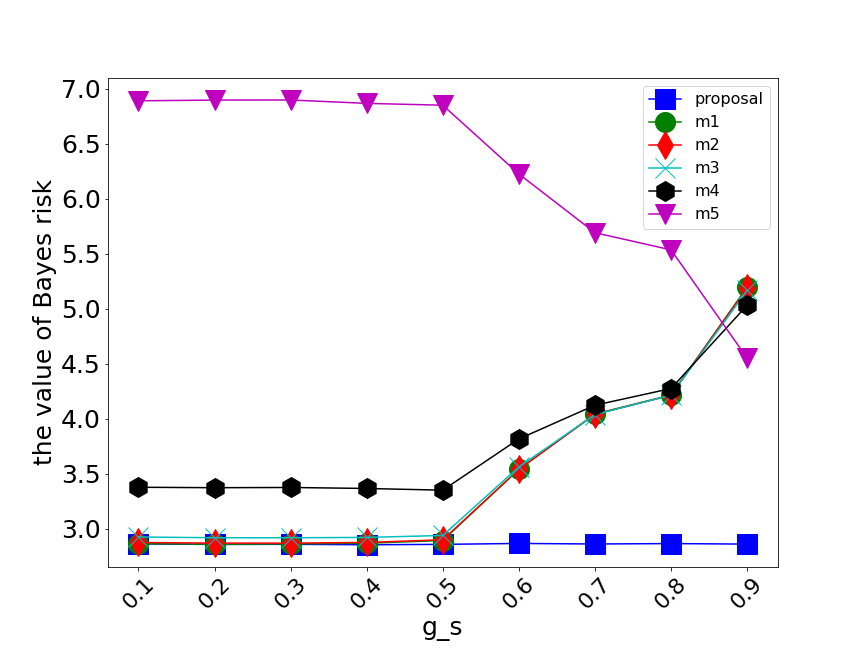}
\caption{Value of Bayes risk function $BR(\delta)$ for each method}
\label{experiment2}
\end{figure}

\begin{table}[tbp]
\centering
\caption{Average depth of $m \in \mathcal{M}$ for each $g_s$}
\begin{tabular}{l|c}
$g_s$&Average depth of $m \in \mathcal{M}$\\ \hline\hline
$0.1$&0.000\\
$0.2$&0.290\\ 
$0.3$&0.957\\
$0.4$&0.897\\ 
$0.5$&2.318\\
$0.6$&3.243\\ 
$0.7$&3.974\\
$0.8$&3.732\\ 
$0.9$&4.452\\ \hline
\end{tabular}
\label{experiment2_depth}
\end{table}

The result of this experiment is at Fig. \ref{experiment2}, and TABLE \ref{experiment2_depth} shows the average depth\footnote{The average depth is calculated as $\frac{1}{|\mathcal{L}^{m}|}\sum_{s\in \mathcal{L}^m}i(s)$.}of $m\in \mathcal{M}$ for each $g_s$.

According to this result, we can confirm that the proposed method achieved the minimum value of the Bayes risk function for all noise variances. Moreover, each of $\delta^i$ depends on the depth of the $m \in \mathcal{M}$. For example, the deeper the average depth of the generated trees is, the higher the value of the Bayes risk function of $\delta^1, \delta^2, \delta^3, \delta^4$ is, and vice versa. Therefore, we can confirm the effectiveness of the proposed method.

\section*{Acknowledgment}
The author would like to thank my family for supporting me in my daily life and the members of the Matsushima laboratory for their meaningful discussions.
\appendices

\section{Proof of Theory \ref{theorem_posterior}}\label{proof_posterior}
In the following, let $C \in \mathbb{R}$ be a constant. First, we show the equation (\ref{p(x|m,y)}).
\begin{align}
&\ln p(\bm{x}|m,\bm{y})\\
=&\ln p(\bm{y}|\bm{x},m)+\ln p(\bm{x}|m) +C \label{lnp(x|my)}\\
=&\ln \mathcal{N}(\bm{y}|\bm{x}, \sigma^2_{\bm{\epsilon}}I)+\ln \mathcal{N}(\bm{x}|(W^m)^{\top}\bm{\mu}^m, \sigma^2I)+C  \\
=&-\frac{1}{2\sigma^2_{\bm{\epsilon}}}(\bm{y}-\bm{x})^{\top}(\bm{y}-\bm{x})\nonumber\\
&-\frac{1}{2\sigma^2}(\bm{x}-(W^m)^{\top}\bm{\mu}^m)^{\top}(\bm{x}-(W^m)^{\top}\bm{\mu}^m)+C \\
=&-\frac{1}{2}\bigg(\frac{1}{\sigma^2}+\frac{1}{\sigma^2_{\bm{\epsilon}}}\bigg)\bm{x}^{\top}\bm{x}\nonumber\\
&+\bigg(\frac{1}{\sigma^2_{\bm{\epsilon}}}\bm{y}+\frac{1}{\sigma^2}(W^m)^{\top}\bm{\mu}^m\bigg)^{\top}\bm{x} +C.
\end{align}
Let us define $\tilde{\sigma}^2,\tilde{\bm{\mu}}^m$ as follows.
\begin{align}
\tilde{\sigma}^2&\coloneqq \frac{\sigma^2\sigma^2_{\bm{\epsilon}}}{\sigma^2+\sigma^2_{\bm{\epsilon}}},\\
\tilde{\bm{\mu}}^m &\coloneqq \tilde{\sigma}^2\bigg(\frac{1}{\sigma^2_{\bm{\epsilon}}}\bm{y}+\frac{1}{\sigma^2}(W^m)^{\top}\bm{\mu}^m\bigg).
\end{align}
Therefore, 
\begin{align}
&\ln p(\bm{x}|m,\bm{y})\\
=&-\frac{1}{2\tilde{\sigma}^2}(\bm{x}-\tilde{\bm{\mu}}^m)^{\top}(\bm{x}-\tilde{\bm{\mu}}^m)+C \\
=&\ln \mathcal{N}(\bm{x}|\tilde{\bm{\mu}}^m, \tilde{\sigma}^2 I).
\end{align}
Next, we show the equation (\ref{p(m|y)}).
\begin{align}
&\ln p(m|\bm{y})\\
=&\ln p(\bm{y}|m)+\ln p(m)+C\\
=&\ln \int p(\bm{y}|\bm{x},m)p(\bm{x}|m) \mathrm{d}\bm{x}+\ln p(m)+C\\
=&\ln \int \mathcal{N}(\bm{y}|\bm{x},\sigma^2_{\bm{\epsilon}}I)\mathcal{N}(\bm{x}|(W^m)^{\top}\bm{\mu}^m,\sigma^2I) \mathrm{d}\bm{x}\nonumber\\
&+\ln p(m)+C\\
=&\ln \mathcal{N}(\bm{y}|(W^m)^{\top}\bm{\mu}^m,(\sigma^2+\sigma^2_{\bm{\epsilon}})I)+\ln p(m)+C\\
=&-\frac{1}{2(\sigma^2+\sigma^2_{\bm{\epsilon}})}(\bm{y}-(W^m)^{\top}\bm{\mu}^m)^{\top}(\bm{y}-(W^m)^{\top}\bm{\mu}^m) \nonumber\\
&+\ln p(m)+C\\ 
=&-\frac{1}{2(\sigma^2+\sigma^2_{\bm{\epsilon}})}(\bm{y}^{\top}\bm{y}-2\bm{y}^{\top}(W^m)^{\top}\bm{\mu}^m+(\bm{\mu}^m)^{\top}\bm{\mu}^m) \nonumber\\
&+\ln p(m)+C\\
=&\frac{1}{(\sigma^2+\sigma^2_{\bm{\epsilon}})}\bigg(W^m\bm{y}-\frac{\bm{\mu}^m}{2}\bigg)^{\top}\bm{\mu}^m +\ln p(m)+C\\
=&\frac{1}{(\sigma^2+\sigma^2_{\bm{\epsilon}})}\bigg\{\sum_{s\in \mathcal{L}^m}\bigg(W_s\bm{y}-\frac{\bm{\mu}_s}{2}\bigg)\bigg\}^{\top}\bigg(\sum_{s'\in \mathcal{L}^m}\bm{\mu}_{s'}\bigg)\nonumber\\
&+\ln p(m)+C\\
=&\frac{1}{(\sigma^2+\sigma^2_{\bm{\epsilon}})}\sum_{s\in \mathcal{L}^m}\bigg(W_s\bm{y}-\frac{\bm{\mu}_s}{2}\bigg)^{\top}\bm{\mu}_s+\ln p(m)+C \\
&\bigg(\because s\neq s'\Rightarrow \bigg(W_s\bm{y}-\frac{\bm{\mu}_s}{2}\bigg)^{\top}\bm{\mu}_{s'}=0\bigg) \nonumber\\
=&\log \prod_{s\in \mathcal{L}^m}\psi_s(1-g_{s})\prod_{s\in \mathcal{I}^m} g_{s'}+C.
\end{align}
Let us denote $\ln \psi_s$ as follows.
\begin{align}
\ln \psi_s &\coloneqq \frac{1}{\sigma^2+\sigma^2_{\bm{\epsilon}}}\bigg(W_s\bm{y}-\frac{\bm{\mu}_s}{2}\bigg)^{\top}\bm{\mu}_s.
\end{align}

We assume the following lemma, which will be proved later.

\begin{lemma}\label{lemma_posterior}
Let us denote $\tilde{\psi}_s,\tilde{g}_s$ as follows.
\begin{align}
\tilde{\psi}_s &\coloneqq
\begin{cases}
\psi_s &(s\in \mathcal{L})\\
(1-g_s)\psi_s+g_s\prod_{s'\in \text{Ch}(s)}\tilde{\psi}_{s'} &(\text{otherwise}),
\end{cases}\\
\tilde{g}_s &\coloneqq
\begin{cases}
g_s &(s\in \mathcal{L})\\
\frac{g_s\prod_{s'\in \text{Ch}(s)}\tilde{\psi}_{s'}}{\tilde{\psi}_s}&(\text{otherwise}).
\end{cases}\label{tilde_g}
\end{align}
In this case, the following equation holds.
\begin{align}
\prod_{s\in \mathcal{L}^m}(1-\tilde{g}_s)\prod_{s'\in \mathcal{I}^m}\tilde{g}_{s'}=\frac{1}{\tilde{\psi}_{s_{\text{r}}}}\prod_{s\in \mathcal{L}^m}\psi_s(1-g_{s})\prod_{s\in \mathcal{I}^m} g_{s'}.
\end{align}
Because $\tilde{g}_s$ is in the range of $[0,1]$, we can show the following equation in the same way for the equation $\sum_{m\in \mathcal{M}}p(m)=1$.
\begin{align}
\sum_{m\in \mathcal{M}}\prod_{s\in \mathcal{L}^m}(1-\tilde{g}_{s})\prod_{s'\in \mathcal{I}^m}\tilde{g}_{s'}=1.
\end{align}
\end{lemma}

Using the above lemma, the following equation can be derived by setting $C=-\ln \tilde{\psi}_{s_{\text{r}}}$ 
\begin{align}
&\ln p(m|\bm{y})=\ln \prod_{s\in \mathcal{L}^m}(1-\tilde{g}_s)\prod_{s'\in \mathcal{I}^m}\tilde{g}_{s'}.
\end{align}
\section{Proof of Lemma \ref{lemma_posterior}}
We show the proof of Lemma \ref{lemma_posterior}.
\begin{align}
&p(m|\bm{y})\\
=&\prod_{s\in \mathcal{L}^m}(1-\tilde{g}_s)\prod_{s'\in \mathcal{I}^m}\tilde{g}_{s'}\\
=&\prod_{s\in \mathcal{L}^m\cap \mathcal{L}}(1-\tilde{g}_s)\prod_{s'\in \mathcal{L}^m\backslash \mathcal{L}}(1-\tilde{g}_{s'})\prod_{s''\in \mathcal{I}^m}\tilde{g}_{s''}\\
=&\prod_{s\in \mathcal{L}^m\cap \mathcal{L}}(1-g_{s})\prod_{s'\in \mathcal{L}^m\backslash \mathcal{L}}\psi_{s'}(1-g_{s'})\nonumber\\
&\times \prod_{s''\in \mathcal{I}^m}\frac{g_{s''}\prod_{s'''\in \text{Ch}(s'')}\tilde{\psi}_{s'''}}{\tilde{\psi}_{s'''}}\quad \because (\ref{tilde_g})\label{tmp_lemma1}\\
=&\frac{1}{\tilde{\psi}_{s_{\text{r}}}}\prod_{s\in \mathcal{L}^m\cap \mathcal{L}}(1-g_{s})\prod_{s'\in \mathcal{L}^m\backslash \mathcal{L}}\psi_{s'}(1-g_{s'})\nonumber\\
&\times \prod_{s''\in \mathcal{I}^m}g_{s''}\prod_{s'''\in \mathcal{L}^m\cap \mathcal{L} }\tilde{\psi}_{s'''}.\label{tmp}
\end{align}
In (\ref{tmp_lemma1}), $\tilde{\psi}_{s}(s\in \mathcal{S}\backslash(\{s_{\text{r}}\}\cup (\mathcal{L}\cap \mathcal{L}^m))$ appeared once in the denominator and numerator. Hence, only $(\tilde{\psi}_{s_{\text{r}}})^{-1}$ remains.

Therefore,
\begin{align}
(\ref{tmp})=&\frac{1}{\tilde{\psi}_{s_{\text{r}}}}\prod_{s\in \mathcal{L}^m\cap \mathcal{L}}(1-g_{s})\prod_{s'\in \mathcal{L}^m\backslash \mathcal{L}}\psi_{s'}(1-g_{s'})\nonumber\\
&\times\prod_{s''\in \mathcal{I}^m}g_{s''}\prod_{s'''\in \mathcal{L}^m\cap \mathcal{L} }\psi_{s'''}\\
=&\frac{1}{\tilde{\psi}_{s_{\text{r}}}}\prod_{s\in \mathcal{L}^m}\psi_{s}(1-g_{s})\prod_{s'\in \mathcal{I}^m}g_{s'}.
\end{align}

\section{Proof of Theorem \ref{theorem_recursive}}\label{proof_recursive}
\begin{align}
&\sum_{m\in \mathcal{M}}p(m|\bm{y})(W^m)^{\top}\bm{\mu}^m\label{ex_wmu}\\
=&\sum_{m\in \mathcal{M}}\sum_{s\in \mathcal{L}^m}p(m|\bm{y})W^{\top}_s\bm{\mu}_s\\
=&\sum_{s\in \mathcal{S}}\bigg\{\sum_{m\in \{m'\in \mathcal{M}|s\in \mathcal{L}^m\}}p(m|\bm{y})\bigg\}W^{\top}_s\bm{\mu}_s\\
=&\sum_{s\in \mathcal{S}}\bigg\{(1-\tilde{g}_s)\prod_{s'\in \text{An}(s)}\tilde{g}_{s'}\bigg\}W^{\top}_s\bm{\mu}_s\\
&(\because \text{\cite{nakahara} Theorem 2} )\nonumber \\
=&(1-\tilde{g}_{s_{\text{r}}})W^{\top}_s\bm{\mu}_s+\tilde{g}_{s_{\text{r}}}\nonumber \\&\times\sum_{s\in \mathcal{S}\backslash\{s_{\text{r}}\}}(1-\tilde{g}_s)W^{\top}_s\bm{\mu}_s\prod_{s'\in \text{An}(s)\backslash\{s_{\text{r}}\}}\tilde{g}_{s'}.\label{tmp1_theorem2}
\end{align}
Note that $\text{An}(s)$ is the set of ancestor nodes of $s\in \mathcal{S}$.
\begin{align}
&\sum_{s\in \mathcal{S}\backslash\{s_{\text{r}}\}}(1-\tilde{g}_s)W^{\top}_s\bm{\mu}_s\prod_{s'\in \text{An}(s)\backslash\{s_{\text{r}}\}}\tilde{g}_{s'} \nonumber\\
&=\sum_{s\in \text{Ch}(s_{\text{r}})}(1-\tilde{g}_s)W^{\top}_s\bm{\mu}_s\prod_{s'\in \text{An}(s)\backslash\{s_{\text{r}}\}}\tilde{g}_{s'}+\\
&\sum_{s''\in \mathcal{S}\backslash(\{s_{\text{r}}\}\cup \text{Ch}(s_{\text{r}}))}(1-\tilde{g}_{s''})W^{\top}_{s''}\bm{\mu}_{s''} \prod_{s'''\in \text{An}(s'')\backslash\{s_{\text{r}}\}}\tilde{g}_{s'''}\nonumber\\
&=\sum_{s\in \text{Ch}(s_{\text{r}})}(1-\tilde{g}_s)W^{\top}_s\bm{\mu}_s+ \\
&\sum_{s'\in \mathcal{S}\backslash(\{s_{\text{r}}\}\cup \text{Ch}(s_{\text{r}}))}(1-\tilde{g}_{s'})W^{\top}_{s'}\bm{\mu}_{s'}\prod_{s''\in \text{An}(s')\backslash\{s_{\text{r}}\}}\tilde{g}_{s''} \nonumber\\
&=\sum_{s\in \text{Ch}(s_{\text{r}})}\bigg\{(1-\tilde{g}_s)W^{\top}_s\bm{\mu}_s+ \tilde{g}_{s}\sum_{s'\in \mathcal{S}\backslash(\{s_{\text{r}}\}\cup \text{Ch}(s_{\text{r}}))}\nonumber\\
&(1-\tilde{g}_{s'})W^{\top}_{s'}\bm{\mu}_{s'}\prod_{s''\in \text{An}(s')\backslash(\{s_{\text{r}}\}\cup \text{Ch}(s_{\text{r}}))}\tilde{g}_{s''}\bigg\}.\label{tmp2_theorem2}
\end{align}
Therefore, from (\ref{tmp1_theorem2}) and (\ref{tmp2_theorem2})
\begin{align}
(\ref{ex_wmu})
=&\bigg\{(1-\tilde{g}_{s_{\text{r}}})W^{\top}_s\bm{\mu}_s+\tilde{g}_{s_{\text{r}}}\nonumber \\
&\sum_{s\in \mathcal{S}\backslash\{s_{\text{r}}\}}(1-\tilde{g}_s)W^{\top}_s\bm{\mu}_s\prod_{s'\in \text{An}(s)\backslash\{s_{\text{r}}\}}\tilde{g}_{s'}\bigg\}\label{before}\\
=&(1-\tilde{g}_{s_{\text{r}}})W^{\top}_s\bm{\mu}_s+\tilde{g}_{s_{\text{r}}}\nonumber \\
&\times \sum_{s\in \text{Ch}(s_{\text{r}})}\bigg\{(1-\tilde{g}_s)W^{\top}_s\bm{\mu}_s+ \tilde{g}_{s}\sum_{s'\in \mathcal{S}\backslash(\{s_{\text{r}}\}\cup \text{Ch}(s_{\text{r}}))}\nonumber\\
&(1-\tilde{g}_{s'})W^{\top}_{s'}\bm{\mu}_{s'}\prod_{s''\in \text{An}(s')\backslash(\{s_{\text{r}}\}\cup \text{Ch}(s_{\text{r}}))}\tilde{g}_{s''}\bigg\}.\label{after}
\end{align}
The $\{\}$ in the (\ref{before}) and the $\{\}$ in the (\ref{after}) have the same structure. The same operation can be recursively computed by setting $r_s$ as in equation (\ref{r_s}).

\bibliographystyle{IEEEtran}
\bibliography{IEEEabrv,ref}
\end{document}